# Bioprotectant glassforming solutions confined in porous silicon nanocapillaries

Rémi Busselez[1], Ronan Lefort[1], Qing Ji[1], Régis Guégan[2], Gilbert Chahine[1,3], Mohammed Guendouz[4], Jean-Marc Zanotti[3], Bernhard Frick[5], and D. Morineau[1]

[1]Institut de Physique de Rennes, CNRS-Université de Rennes 1, Rennes, France
[2]Institut des Sciences de la Terre, Orléans, France
[3]Laboratoire Léon Brillouin, CEA-Saclay, Gif-sur-Yvette, France
[4]Laboratoire d'Optronique, Université de Rennes 1, Lannion, France
[5]Institut Laue-Langevin, Grenoble, France

**ABSTRACT**

Glycerol and trehalose-glycerol binary solutions are glass-forming liquids with remarkable bioprotectant properties. In this paper, we address the effects of confining of these solutions in straight channels of diameter $D=8$ nm formed by porous silicon. Neutron diffraction and incoherent quasielastic neutron scattering are used to reveal the different effects of nanoconfinement and addition of trehalose on the intermolecular structure and molecular dynamics of the liquid and glassy phases, on a nanosecond timescale.

**INTRODUCTION**

The manipulation of fluids in nanochannels has become a crucial issue for many foreseen applications in advanced nanomaterials and biotechnology [1]. Fundamental questions arise from the unexpected behavior of fluids confined in capillaries of nanometric dimension, which rule out the validity of some approaches derived from the physics of liquids at the macro or mesoscopic scale. Intensive experimental studies of molecular liquids have shown that confinement on a nanometric scale considerably modifies the structure, phase behavior and molecular dynamics. Recently, much effort has focused on the unusual dynamical properties of low-molecular weight liquids and glassforming systems in mesoporous solids. It reveals a complex entanglement of low dimensionality, finite size and surface effects [2]. In this field, a current challenge is to extend the knowledge of nanoconfined liquids to more complex fluids such as soft-matter or solutions of biological interest. This study focuses on pure and binary solutions comprising two glassforming components (trehalose $T_g=393$ K and glycerol $T_g=190$ K), which exhibit remarkable bioprotectant properties closely related to the nature of their H-bonds structure and glassy dynamics. The aim of the present contribution is to report on the effect of confinement in silicon nanochannels on their structure and microscopic dynamics.

**EXPERIMENT**

Confinement was performed in the nanochannels of the so-called 'columnar form' of PSi [3,4]. These samples were electrochemically etched with a current density of 50 mA.cm$^{-2}$ in a solution composed of HF, H$_2$O and ethanol (2:1:2) according to previous studies [5]. It leads to highly anisotropic porous layers consisting of macroscopically long channels with pore diameter $D\approx 8$ nm running perpendicular to the wafer surface. Scanning electron microscopy (SEM) measurements provide interesting information about the shape of the porosity. It forms a parallel arrangement of not-connected channels of length 200 $\mu$m as shown in figure 1. A noticeable feature of PSi, which is at variance to other unidirectional pores such as MCM-41 or SBA-15, is the rough dendritic structure of the inner pore surface [6]. The latter has been recently shown to introduce strong disorder effects, which influence phase transitions such as capillary condensation, nematic and smectic transitions [7,8] as well as the molecular dynamics [9].

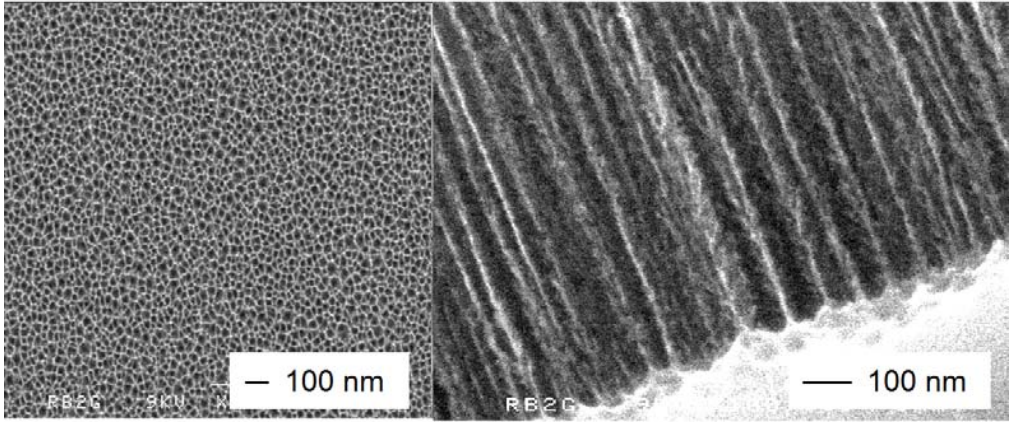

**Figure 1.** Scanning electron micrographs of the porous silicon (a) top view and (b) side view.

We used fully hydrogenated glycerol ($C_3O_3H_8$, Sigma-Aldrich 99%) and anhydrous trehalose ($C_{12}H_{22}O_{11}$, Merck 99%) for the incoherent neutron backscattering experiments (dynamics). For the neutron diffraction experiments (structure), we increased the coherent scattering cross section of the samples by using fully deuterated glycerol ($C_3O_3D_8$, Eurisotop) and partly deuterated trehalose ($C_{12}H_{14}O_{11}D_8$). Partial deuteration of trehalose was achieved by three cycles of chemical substitution of the exchangeable hydrogens of anhydrous trehalose with deuterium oxide followed by lyophilisation.

Glycerol-trehalose binary solutions (80%:20% Wt) were prepared by adding the appropriate mass of crystalline trehalose to glycerol. Just before the execution of neutron scattering experiments, pure glycerol and trehalose-glycerol were filled in PSi samples by capillary impregnation from the liquid phase in a vacuum chamber. The complete loading was achieved under vapor pressure and at a temperature of 333 K, well above melting temperatures. The filled porous membranes were taken out from the vacuum chamber, the excess of liquid was removed from the wafer surface by wiping the samples with filtration papers and the samples were placed immediately in hermetically closed neutron scattering aluminum cells.

The structure of the bulk and confined phases was characterized by the measure of the coherent neutron static structure factor $S(Q)$. The neutron scattering experiments were performed on the double-axis spectrometer G6.1 at the Laboratoire Léon Brillouin neutron source facility (CEA-CNRS, Saclay, France) using a monochromatic incident wavelength of 4.7 Å, which allows covering a range of transfer of momentum from $Q=0.13$ Å$^{-1}$ to $Q=1.8$ Å$^{-1}$.

The molecular dynamics was characterized by the measure of the incoherent elastic structure factor $S(Q,\omega\approx 0)$. The quasielastic neutron scattering experiments were carried out using the high resolution back scattering spectrometer IN16 at the Institut Laue Langevin (Grenoble, France). A standard configuration of the IN16 spectrometer was chosen with Si(111) monochromator and analysers, which corresponds to an incident wavelength of 6.271 Å and results in an energy resolution (FWHM) of 0.9 $\mu$eV with a range of transfer of momentum between $Q=0.2$ Å$^{-1}$ and $Q=1.9$ Å$^{-1}$.

## RESULTS AND DISCUSSION

### Static structure factors

The static structure factor $S(Q)$ of the bulk liquid glycerol at $T=300$ K is shown in solid line in figure 2(a). The range of transfer of momentum covered by this experiment allows focusing on the region of the main diffraction peak (around $Q=1.4$ Å$^{-1}$ for glycerol). In disordered phases (liquid or glass), the region of the main diffraction peak mostly reflects the contributions from the intermolecular spatial correlations to the total static structure factor. It is well-known that the intermolecular correlations progressively vanish at higher $Q$ and that the structure factor is essentially defined by the intramolecular form factor for $Q$ values typically larger than 4 Å$^{-1}$ [10]. Conversely, supramolecular long range fluctuations dominate the intensity of the structure factor in the limit of low $Q$ (i.e. $Q<0.1$ Å$^{-1}$). Hence, in the case of the pure liquid glycerol in bulk conditions, the structure factor is related to the static compressibility in the limit of $Q\sim0$.

At lower temperature, $T=130$ K, the structure factor of pure glycerol remains similar to the liquid one. This is typical for an amorphous solid, which retains the disordered character of the liquid phase below the glass transition temperature ($T_g=190$ K). One notices a shift to high $Q$ of the position of the main diffraction peak (of about 4 %), which results from the increase of the microscopic packing fraction due to the increase of the average density at low temperature.

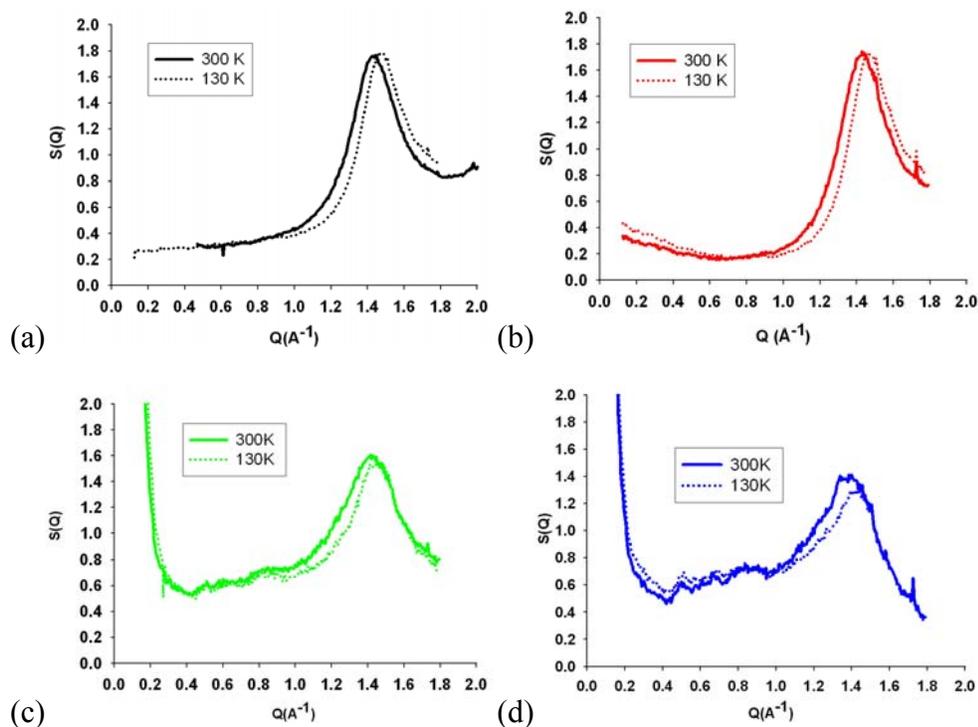

(a) (b)
(c) (d)

**Figure 2**. Static structure factors of glycerol and the glycerol-trehalose solution in bulk and confined in porous silicon layers measured in the liquid state at $T=300$ K (solid line) and in the glass at $T=300$ K (dashed line). (a) bulk glycerol, (b) bulk glycerol-trehalose solution, (c) confined glycerol and (d) confined glycerol-trehalose solution.

The static structure factor of the bulk binary trehalose/glycerol solution are shown in figure 2(b) at the same temperatures $T$=300 K and $T$=130 K. At variance to pure liquid glycerol, a significant increase of intensity is noticeable in the limit of low $Q$ (i.e. $Q$<0.7 Å$^{-1}$). This feature is a direct consequence of supramolecular inhomogeneities in the binary solution, which remains however homogeneous at the macroscopic scale. This suggests the formation of intermediate range clustering effects, probably related to H-bond interactions and concentrations fluctuations in the binary solution. At $T$=130 K, the main diffraction peak is shifted to high $Q$ similarly to the case of pure glycerol. Additionally, the low $Q$ excess intensity rises, which illustrates the increase of the heterogeneous nature of the binary solution mesoscopic structure.

The static structure factors of the confined systems are shown in figure 2(c) and 2(d) for the pure glycerol and trehalose/glycerol solution, respectively. Despite the use of a much smaller amount of materials than for the bulk systems, resulting in a lower signal-to-noise ratio, one can notice three important features. First, one notices that the structure factors still present broad peaks, even at $T$=130 K, indicting that the systems are again essentially amorphous. One concludes that confinement in PSi has not induced nucleation and crystallization, the systems remaining either liquids or vitreous on the studied temperature range. Second, there is a sharp increase of intensity at low $Q$, which is ascribed to the mesoscale structure of the nanoconfined system (i.e. solid-fluid interface, pore form factor and pore-pore correlations). The low $Q$ excess intensity ascribed to concentration fluctuations in the case of the bulk binary solution is obviously overwhelmed by this contribution for the confined system. A last feature concerns the main diffraction peak itself, which is systematically broader and shifted to lower $Q$ in confinement with respect to the bulk cases. These changes of the shape of the main different peak may have different origins, which are probably concomitant. A broadening of the structure factor is usually attributed to the decrease of the static correlation length, which characterizes the propagation of short-range order to the intermediate range. Indeed, ongoing molecular simulations of liquid glycerol in porous silicates have revealed a disruption of the H-bonds network by confinement [11]. A reduction of the fluid density could induce a shift of the main diffraction peak to low $Q$. There are indeed a number of experimental evidences that confined fluids can have a lower density than the bulk ones [12]. A recent positron annihilation lifetime spectroscopy study of glycerol confined in 7 nm silica glass has also concluded to a higher free volume [13]. Quantifying the extent of this confinement effect on the statics is however delicate unless contributions from fluid-matrix cross correlations and topological excluded volume effects are fully accounted. Indeed, for other fluids confined in smaller pores, these effects have been shown to contribute significantly to an observed shift and broadening of the main diffraction peak [14].

**Dynamics probed by elastic fixed window scans**

In order to monitor the molecular dynamics from the liquid to the glassy state, elastic fixed window scans (EFWS) have been acquired during cooling from 350 K to 10 K, as shown in figure 3. In an EFWS, the measured intensity is in principle $S_{inc}(Q,\omega=0)$. It actually integrates all purely elastic or quasielastic contributions slower than the energy resolution defining the "fixed window" (i.e. 0.9 $\mu$eV FWHM). At high temperature (above 350 K), the elastic intensity is below the detection level for bulk and confined glycerol and extremely small for the bulk and confined binary solution. This result is expected for a liquid for which translational diffusion completes within the experimental timescale, giving rise to a quasielastic line that fully exceeds the elastic resolution. The elastic intensity gradually increases on cooling to reach a value of about 0.8 at 200 K. This feature is common to glass-forming systems. It most probably reflects the progressive slowdown of the translation diffusion and possibly rotational molecular dynamics, which leads to a sharpening of the associated quasielastic scattering. The contribution of this quasielastic scattering to the fixed energy window intensity increases until the quasielastic lines are not broad enough to be discriminated from a true elastic peak within the experimental resolution. Below this temperature of onset of structural relaxation processes, only localized degrees of freedom of the glass phase (conformation fluctuations and Debye-Waller factor, comprising inter- and intramolecular vibrational modes) result in a remaining (but weaker) temperature dependence of the elastic component.

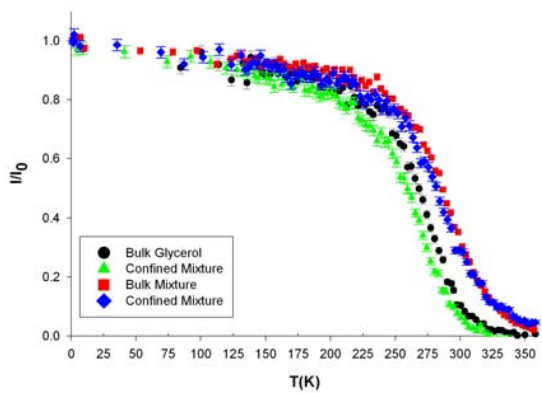

**Figure 3**. Fixed window elastic scattering intensity of glycerol and the glycerol-trehalose solution in bulk and confined in porous silicon layer obtained during a scan on cooling for transfer of momentum $Q$=1.33 Å$^{-1}$.

A comparison between the EFWS displayed in figure 3 for the four different systems reveals some distinct effects on the relaxation dynamics of glycerol solutions on the nanosecond timescale of both confinement and of the addition of trehalose. The elastic intensity of the bulk trehalose-glycerol solution qualitatively resembles the bulk glycerol one, except for an overall shift to higher temperature of about 20 K, when measured at half the maximum elastic intensity. This overall slowing down of the relaxation dynamics of the binary solution on increasing the amount of trehalose is mostly related to the change in the glass transition temperature.

Conversely, confinement leads to an apparent shift to lower temperature of the elastic intensity of about $\Delta T$= -7 ±2 K when measured a half maximum intensity. This feature is observed for pure glycerol and for the binary solution. It signifies that the slowdown on cooling of the microscopic molecular dynamics as probed on the nanosecond timescale is reduced in PSi nanoporous channels. It reveals a significant effect of confinement on the glassy dynamics of these bioprotectant solutions.

**CONCLUSIONS**

We have shown that confinement in porous nanochannel of silicon ($D$=8 nm) affects the structure factor of glycerol and trehalose-glycerol bioprotectant solutions in the liquid and glassy states. It corresponds essentially to a broadening and a shift to high $Q$ of the main diffraction peak. Moreover, mesoscopic concentration fluctuations are revealed in the binary solution. They lead to an excess of intensity at low $Q$ for the bulk system, which is overwhelmed in confinement by additional contributions from the mesostructured porous system. The molecular dynamics probed by incoherent elastic fixed window scans is significantly slowed down by the addition of 20% of the bioprotectant trehalose in glycerol, whereas confinement leads to an apparent acceleration of the dynamics. It is suspected that changes in the static properties of the confined systems could partly contribute to the observed changes of the molecular dynamics. Indeed, both the frustration of the H-bonds network and a lower density induced by confinement should directly affect the molecular dynamics of the confined fluids.

**ACKNOWLEDGMENTS**

This study was part of the PhD thesis of R. B., supported by a doctoral research grant from the *Brittany Region*. Financial supports from the *Centre de Compétence C'Nano Nord-Ouest* and *Rennes Metropole* are expressly acknowledged.